# Skeletal Structures of the Ocean, Hypotheses and Interpretation of the Phenomenon

V. A. Rantsev-Kartinov

INF RRC "Kurchatov Institute", Kurchatov Sq. 1, Moscow, 123182, Russia,
Tel.: 7 (095) 196 7334, Fax: 7 (095) 943 0073, E-mail: rank@nfi.kiae.ru

## 1. Introduction.

An analysis of a photographic databases of an ocean surface (OS) images (taken from various altitudes and for various types of the rough OS) has leaded of the author to observation of an ocean skeletal structures (OSS) [1]. The topology of the OSS appears to be identical to that of a skeletal structures (SS) which have been formerly found earlier in a wide range of length scales, media and for various phenomena [2-4]. In paper [2c,d,f] it was described in detail and given an explanation of same properties of such SS and suggested hypotheses, the main idea of which is *only microdust component with its quantum connections may be responsible for the SS, which observed in plasma* [2c] (this hypothesis has been suggested by my co-author under the given publications). The total papers [2e,3a] which is concerning to the SS gives a conclusion about of role of the nanodust in formation of the SS to a very wide range of length scales which have covered [1e, 2a] as much as about 30 orders of magnitude.

## 2. Hypotheses.

In the paper [1] it was suggested hypotheses about formation of the OSS. The basic theirs states are: **a**) – a structure - generating dust, which delivers in atmosphere by the volcanic activity, underlies of the OSS (as well as in base of the SS, which has been described earlier [2-4]), when a carbon nanotubs (CNT) or similar particles from another chemical elements may to be a basic elements; **b**) – under influence of the Earth electrical field and the atmospheric electricity such dust is able to draw up the SS of a clouds whose fragments falls on the OS due to a different atmospheric phenomena; **c**) – the SS have very active surface and without zero buoyancy, therefore they floats up to the OS is adsorbing the dissolved in the water air on its own surface and partially is filling in by a sea foam; **d**) – the SS (it is creating on its own surface of three substances separation (solid, fluid and gas)) supports of the possibility of the surface tension forces (STF) action (due to incomplete wetting ) even under the OS and links of a separate blocks into the single whole OSS; **e**) – the OSS and its separate blocks strength is determined by a compactness of the SS (which ware fell out on the OS) packing by its own fragments, they are, a blocks of a previous generations; **f**) – the OSS skeletal characteristics becomes apparent the fact that the individual straight and rather strong its blocks may be joined flexibly (due to the action of STF), similarly to joints in a skeleton.

## **3 Observations of the OSS.**

In [1] it gives a short list of a type blocks (which is observed in the OSS), some short descriptions and fragments of some images of them. One in a conclusion of [1] is the fact of increasing of a basic sizes of the observing OSS blocks during increasing of the rough OS level. In this paper it gives estimations of a cohesion strength and buoyancy of the OSS blocks in the form of a vertically oriented floating cylinders (VFC). They are a coaxial – tubular (CT) blocks closely packed (during a storms) by the SS fragments (which are fell out on the OS) and blocks of the previous generations. In such case they are filled in with closely/partially packed blocks of a smaller size, up to thin capillaries of tens of microns in size (in the form of, e.g., the CNT or similar particles from another chemical elements). Such capillaries fill up of the OSS blocks of anyone types. Constructions of individual blocks (in frames of own topology) may to be able with a different open-working. At that, this open-working of blocks may do not lead to loss its strength. Just like this is the CT block with a structures like "cartwheel" (CW) on its butt-ends and a parallel knitting-needles along of a cylinder lateral surface, and which represents the CT block of a "squirrel-wheel" (SW) type. Such block is very strong, stable as regards to conservation of own form and has lightest weight of its construction. At this everyone power elements of them will may be assembled from mentioned above capillaries. CT block may to have a smooth side surface or fills up by a tubes of smaller diameters, which in one's turn assembled from the CNT capillaries. These capillaries may to partially/closely fills up by the sea water with air (foam) providing thereby some buoyancy of all the OSS as a whole. For demonstration of the VFC it gives an example from [1], which is a fragment of the image of the OS, taken during the Hurricane Belle from 500 feet altitude [5], and represents here as a Fig.1. In order to discover of a clarity of the observing VFC structure here it is applied a method of multilevel dynamical contrasting (MMDC) [2a,b,3a], which has been worked up by author of this paper.

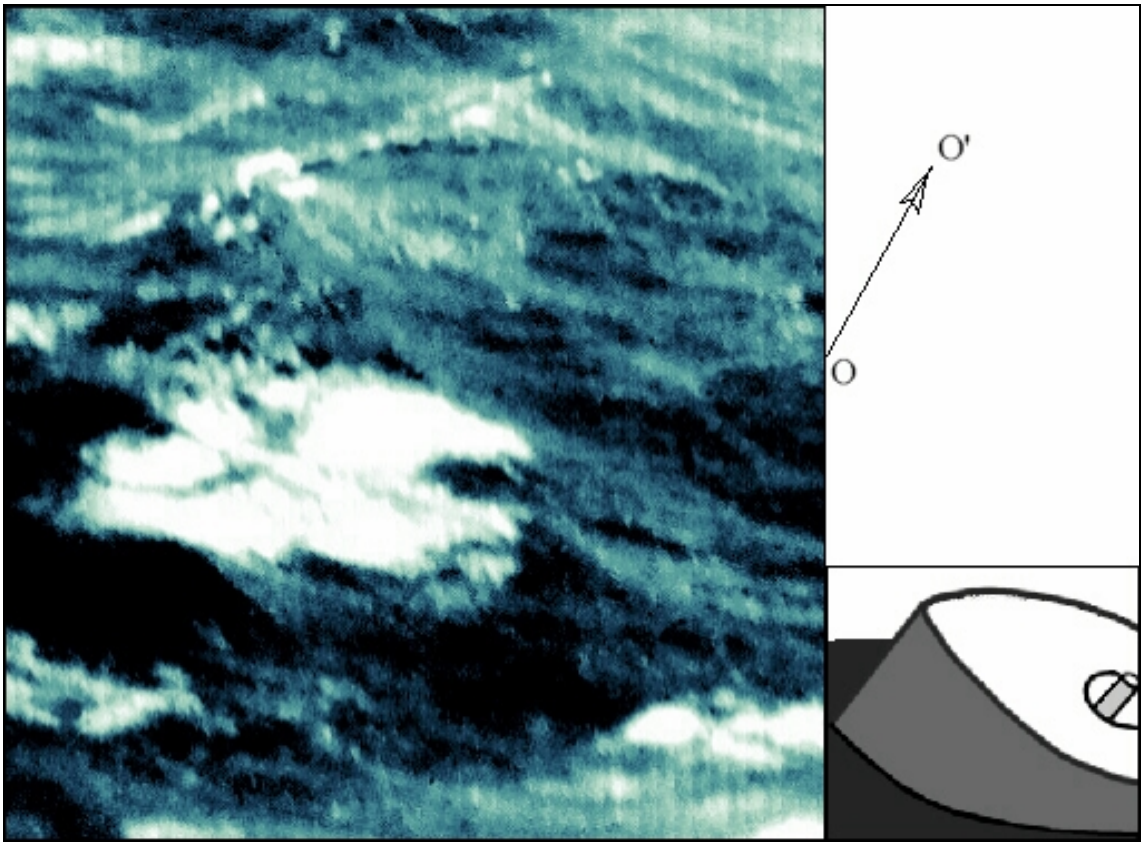

**Fig. 1.** A fragment of the image of the sea surface, taken during the Hurricane Belle from 500 feet altitude (NOAA collection [4]), MMDC. Image's width is ~15 m. One may see, in the front of the image, almost a half of a vertically oriented floating cylinder, VFC, of diameter ~ 25 m. Diameter of central axial tube is ~ 3 m, diameter of a dark rings in the VFC edge is ~1.5 m. The VFC is ~ 1.5 m above the sea's local level. OO' – direction of the VFC axis. An angle between of the O'O direction and vertical line is about 30˚. Below, at the right, in window it is given schematic image of this VFC.

# 4. Interpretation of the observing phenomenon.

The mentioned above hypotheses **d** notes – *presence of medium in a different phase state may to introduce considerable amendments into already considered earlier of processes of formation the SS* [2e, 3a,]. Let us to estimate now: "Is that enough of the STF and capillary phenomena in order to explain of presence of observed into the ocean waves of the VFCs, which is filled up with a sea water together a foam and towers above sea level up to one meter?" Let us assume, that during a storm some forces, which are able to push out of a construction (assembled of the thin capillaries) on a big enough height above sea level are existing, and a capillary walls are moistened. In such case it is possible to estimate of the height of fluid lifting on these capillaries. If the VFC construction filled in with closely packed thin capillaries, which are oriented parallel to the VFC lateral surface, and the water weight of its column is balanced by the STF into these capillaries. In that case there are no forces, which are able to tensile of the block structure, and itself weight is balanced by forces which are responsible for the block buoyancy. Between the capillaries which fillings up of the VFC it is form capillaries with sizes of smaller to basic ones. In such ideal block the water weight lifted above sea-level is balanced by the capillary forces. Such block will to keep its form, and its butt-end surface will be washed by the water due to the rough OS as a solid/open-working surface, according to its own construction. So, if the VFC volume partially fills up by the sea foam (due to the rough OS), then it floats under the middle sea level on some height. The capillaries forces hold the water in its canals. Until the sea foam is holding into the capillary canals the VFC holds oneself floating and conserves of its own geometrical form with the vertical oriented side surface. The filling up of the VFC by the foam is a result of a wave overturning on its butt-end during of a storm, that takes place even in the open ocean. Along the vertical direction the water of the VFC, which the top butt-end is floating over the middle sea level, stretched by its own weight. In the radial direction the VFC structure has the strength characteristic of a cylindrical wisp assembled by the thin parallel capillaries connected between themselves due to the STF. It is ease to show that the corresponding breaking point $E$ for the VFC it is able to estimate as:

$$E \approx 4 \cdot 10^{-6} \, \sigma/d \; \text{kg/cm}^2, \tag{1}$$

where $\sigma$ - a coefficient of the STF of the water [din/cm] and $d$ - a capillarity diameter [cm], which filling up of the VFC body. That is to say, the thin capillaries strengthen the VFC construction.

When the VFC axis is deflected on an angle $\alpha$ relatively of the vertical direction, that the capillary length filled up by the water is:

$$l \propto 1/\cos\alpha .\qquad(2)$$

Then, at inclinations of such block among waves fluid will be to overflow of capillaries of this block, and themselves the block will to sink a little and to lose of a steadiness with partial destruction of its inclined the lateral surface, that is observing sometimes.

Let us now to estimate, what are maximum sizes of diameters of the basic capillaries, which confines of the VFC. If the water is lifted by these capillaries up to high $h$, this estimate follows from an inequality:

$$d \leq 4\sigma/gh\rho ,\qquad(3)$$

where g – acceleration of gravity, $\rho$ - the water density. Substitution here of the book reference data gives the maximum size of these capillaries diameters. If $h$ ~ 1 m, then the capillary limitation is:

$$d \leq 3\cdot 10^{-3}\ \text{cm}.\qquad(4)$$

In the presence of the large enough buoyancy (in order to hold a water column, which is lifted up on some height $h$ over the middle sea level due to the capillary forces) such capillary construction will to hold oneself floating and conserve of its own cylindrical form with the vertical oriented side surface, as a fishing float.

Let us now to estimate of the VFC buoyancy and ability of it to carry of such loading as the water column weight with a diameter equal to an external diameter of a floating structure and height $h$ over the middle sea level. From the beginning for this it is necessary to set oneself a model of building of the self-similar tubular structures. With that end in view one/two-level CNT 100 Å long we will choose as a basic block of our building. Such basic CNT with a large probability obtains under action of any energetic influence on a crystalline carbon. The carbon crystalline plane is a mosaic of a flat rags (with typical size ~ 100 Å and compactly laid dawn with carbon hexahedrons of side ~ 1,42 Å) with a relaxed connexions between them. The CNT of the first generation – it is rolled up in a scroll such rag, which contains ~ 8 $10^3$ of carbon atoms and a mass $m_1$ ~ 1,7 $10^{-19}$ g. Then we will lay dawn some square by the CNTs of firs generation such so they forms the same hexahedrons but with side of ~ 100 Å and will roll up of it into a similar tube of second generation and so on. The mass, length and diameter of such tubes of nth generation is determined in such case correspondingly as:

$$m_n \sim 1{,}7 \cdot 10^{-19} \cdot 10^{4(n-1)} \text{ g}, \quad l_n \sim 1{,}42 \cdot 10^{-8} \cdot (75)^n \text{ cm}$$

$$\text{and } d_n \sim 3 \cdot 10^{-7} \cdot (75)^{n-1} \text{ cm}. \tag{5}$$

From here it is follows in the Earth conditions it would be able to observe the CW structure of the 9-th generation, since at that $d_9 = 3 \cdot 10^3 \ km$. But a maximal size of the CW diameter, which at the present moment (at analysis of a data base, which the author has available, of the OS images obtained from a cosmos orbit altitude) has revealed by the author of this paper in the meantime consists ~ 650 km. On the OS it is able to observe the full-length CT structures (see (5)) as the VFC up to the 7-th generation inclusive. Indeed, $n = 7$ agree with $d_7 \sim 530 \ m$ and $l_7 \sim 2 \ km$. In that way, the CT structure as the VFC of 7-th generation is able to move free along the OS there where it is allowed by the ocean depth. The similar structures of the 8-th generation have already the diameter $d_8 = 40 \cdot km$ and length $l_8 \sim 140 \ km$. Therefore in the open ocean of such structure it would be able to see only as the HFC or as the VFC, but with a shortened length.

The butt-ends of the CNTs up to 3-rd generation may to become clogged (providing thereby the SSO buoyancy) with a flat fish-scale of phytoplankton, which has size about

$$\delta \sim 5 \ 10^{-4} \text{ cm} \tag{6}$$

Between that size and of the capillary diameter, which is able to lift up the water column up to height ~ 1 m, it takes place a gap of order magnitude. Aside from the CNTs can intensively to absorb the air dissolved in the sea water, adsorb it on theirs own surface and (according to formation and gathering of the air bubbles on its surface) create some additional buoyancy to the SSO blocks. It turns out, that only the CNTs of 3-rd generation are able to lift up the water column on the meter height. Indeed,

$$d_3 \sim 1{,}7 \cdot 10^{-3} \text{ cm,} \tag{7}$$

that agrees with condition (4) and they do not plug up by the phytoplankton particles, since ones in some times smaller (see (6)) of the obtained diameter. The specific gravity of selected by us of the SSO block (for example with diameter ~ 8 m) is determined by a number of its generation. This number easy determines (see(5)) and it is equal n ~ 6. From here it is simple to calculate a density of such initial tube skeleton ~ 1,3 $10^{-5}$ ρ, therefore at our estimates by this magnitude may to ignore. During a strong storm the initial SS strength is not enough to hold out against crashes of the running high ocean. As that follows from a visual analysis of the OS images during the strong storms, this skeleton breaks to pieces and forms up a new more strong structure, which is a tubular structure with tubes corresponded to the determined generation, but they are compactly fills in by the tubes many smaller diameter (another boundary case). In which connection, as this is seen from a visual analysis at such filling (by the CT blocks of the previous generations) inside of the VFC it may to remain some of such tubes unfilled in.

Now we will try to describe of modification of the initial SS in the sea water and transformation of its into the OSS. As stated above, we can be satisfied only by the 3-rd generation tubes. Therefore, in the VFC building the very same tubes will be able to take part. If very strong the rough of the OS destroys of same part of the initial SS, so the OS is covered by these tubes (capillaries) as a compact carpet. Since at a strong wind a waves have the primary orientation that such tubes lines up along the wave crest. They install of a connections between themselves due to the STF, becomes longer and covers the OS by the compact carpet. The images analysis shows a length of similar filaments (and of even considerably larger diameter) are able achieve at times of length up to one hundred meters. Intensification of the rough OS leads to rolling up of this carpet into a cylindrical rolls of the compact packed up tubes of smaller diameter at the same time leads to filling in of an escaped destruction blocks of the initial SS by this capillary debris. Diameter of these rolls corresponds to highest possible height of wolves, which are observing at the present moment of time. Statistics of the visual investigation of topology of such observing rolls can to give (in zero approaching) a rule of its construction – the cylinder of present generation is a sheaf of cylinder of a previous generation and (most likely) six the same cylinders around it, which are enclosed into a one cylindrical shell. At that, an unfilled up space which may to take place at such forming can fills in or no by a cylinders (capillaries) of a more early generations. Total percent of filling in may be about 100 %. The mentioned constructing rule of generations in such process is quite natural, because the process of self - assembling and enlargement of cylinders proceeds parallel with process of the intensification of waves and theirs energetic power. At very strong of rough of the OS and overturning of the waves proceeds whipping of the sea foam. It plays an additional part in increasing of the buoyancy of the SS blocks (which forms during that process), as since, it promotes of decreasing of the middle specific gravity of the water, which fills in the blocks bodies of the OSS.

Let us to estimate a specific gravity of carbon in the ideal VFC at assumption such block is completely packed up by the CNT of the 3-rd generation. The estimation gives quantity $\sim 1.3 \cdot 10^{-3} \rho$, that at $\sim 10^2$ times as much than a density of a condensed component into the water of the initial ideal model of the SS. It turns out, at the intensification of the OS rough, near to the water surface the special gravity of a dusty component in the OSS increases. But it makes an insignificant contribution to the middle special gravity of the sea water, because a content of salts and another impurities in it almost at 20 times as much than a part of a discussed us dusty component. It is necessary to note, only at taking a sample of the water inside of such ideal VFC and at a hurricane may to give of such high content of the micro-dust component in the sea water. In any another cases its content downs up to $\sim 1{,}3 \cdot 10^{-5} \rho$. In such cases it is not revealed generally.

Therefore, it is very problematically to reveal its presents into the water by means of simple evaporation and separation. About the its presence into the ocean water it is able to judge only by means of a very delicate physical/chemical analysis or an analysis of a particular properties and phenomena, which becomes apparent at a different of levels of the OS rough, in like manner as in an anomalous atmospheric phenomenon connected with the sea.

Let us now to estimate of the buoyancy of the VFC with the maximal diameter, which was observed, described above and towers above sea-level during some, but finite, time. Assuming that we have convinced oneself of the existence of the VFC and took into account this fact as a grant. From observations it follows that for the similar but a horizontal floating cylinders (HFC), which observes during of the very strong storms and has length up to $L \sim (50\text{-}100)$ m. It is already enough for estimation (of an averaged on the VFC volume) of the specific gravity of the filling in its water in order to it will be able to float up over the sea level.

For the VFC filled in by the water with the middle specific density $\rho_a$ and floated up over the middle sea level on the height $h \sim 1m$, it may to writ down a ratio:

$$\rho/\rho_a = (L-h)/L. \qquad (8)$$

Putting here a numerical values of $L$ and $h$, obtains:

$$\rho_a = (0{,}98 - 0{,}99)\rho. \qquad (9)$$

That is, the total volume of air in the water (which filled in the VFC) consists only (1-2) %. During the time of going out of this air bubbles into an atmosphere and the sea foam from the capillaries the VFC will be to sink a little and down to the water surface. Such anomalous the VFC buoyancy exists due to a difference of two times – the time of going out an air bubbles from the open water and the time of going out of one from the VFC capillaries. The VFC downs to the water surface, when the specific gravity of the water into the capillaries and of the open water have equal. Higher concentration of such bubbles appears when the crest of a covering wave overturns on the VFC or when the sea foam together with the water flows goes into its from below. The last takes place at penetration of a power streams of the foamed water on a high depth and filling up of the VFC during floating of the foamed water to the sea surface. Have imbued with foam the VFC floats on a time. Thus, *the CNT component provides an action of the STF even under the OS creating in that way a force guarantee of a safe keeping of the created OSS blocks and its buoyancy*. The total force of the VFC cohesion (some higher to the OS) may to be very considerable (due to a very developed of a total lateral surface of that block), since it is compactly filled up by the thin interacting capillaries. Since the water level into capillaries is completely determined (see (3)) there are no forces, which would ware able to cut off the VFC near its base. If the VFC assembled of the CT blocks of the 5-th generation, which, in one's turn, in their interior, are filled in with closely packed blocks of a smaller size, up to thin capillaries of tens of micron in size (for example in the form of the CT blocks of the 3-rd generation with diameter $d_3$), it is possible to estimate a minimum angle $\alpha_{min}$ of the VFC declination of the vertical line:

$$\alpha_{min} \approx \arcsin(2\sigma/d\rho g) \qquad (10)$$

If $d = d_5 \sim 10$ cm (see (5)) $\alpha_{min} \approx 1.5 \cdot 10^{-2}$, i.e. such VFC at declination of the vertical line will fall to the water surface because the 5-th generation CT blocks near upper butt-end will tear off from cylinder and bend over the water surface on the side with a negative inclination. The VFC on Fig.1 has an angle of its inclination relatively of the vertical line direction about 30° and it keeps its own geometrical form. From this it follows (see (10)) that

$$d_{max} \leq 4\sigma/\rho g = 3 \; 10^{-1} \text{ cm,} \qquad (11)$$

that almost corresponds to $d_4 \sim 1.3 \; 10^{-1}$ cm. Thus, from observations it follows that:

$$\delta << d_3 \approx d \leq d_4, \text{ that is, } 5 \cdot 10^{-4} cm < d \leq 3 \cdot 10^{-1} cm \qquad (12)$$

During a high levels of the rough ocean surface it strips uncover the HFC structures which are able lengthwise up to 50 – 100 m. At that its aspect ratio may to arrive at value $D/l \sim 3 \cdot 10^{-4}$. During a middle levels of the rough ocean surface such lengthy the HFC sometimes overlap of span between two a neighboring waves as a convex bridge. In such cause it is easy to estimate a middle value of the structure strength. A fragment of the same structure is given in Fig.2.

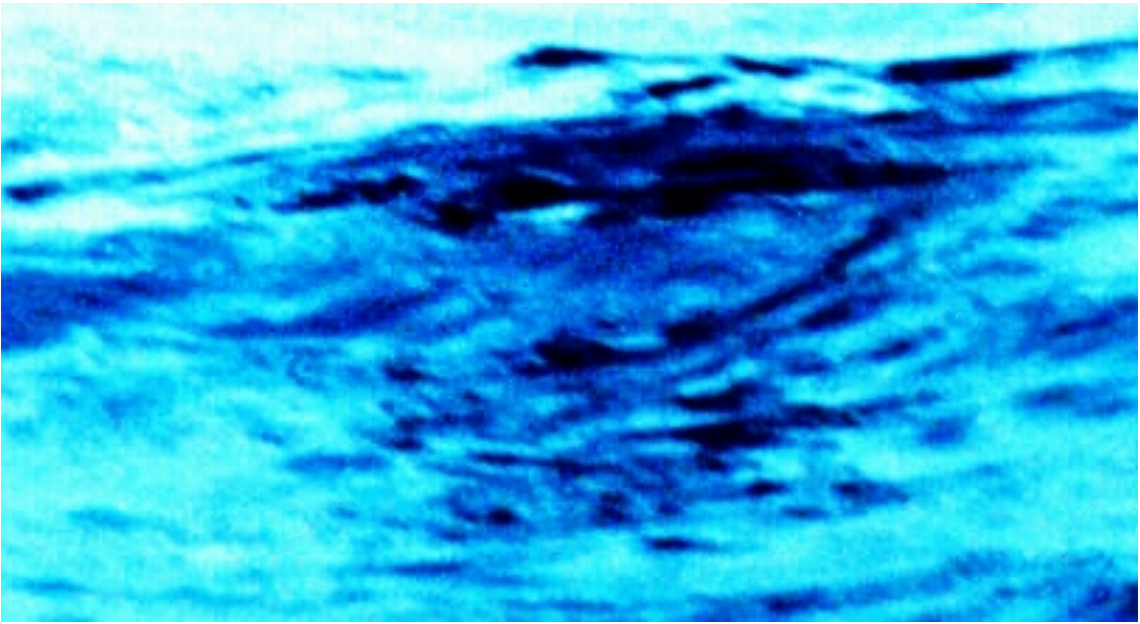

**Fig.2.** Here it is used the MMDC, image's width is ~ 1.7 m. At this point the OSS reveals itself as a long and nearly straight line dendrite construction, branches of which immersed in some depth. This construction overlap of span between two a neighboring waves as a convex bridge.

Supposing a weight of such beam determines by the water specific gravity (which fills its) and to taking into account the fragment scale mentioned above it is able to estimate an average value of the breaking point E for its. It turns out, that $E \geq 0.1$ kg/cm$^2$ ($D \sim 10$ cm and $l \sim 100$ cm). Then, if it will assumed such structure is a bunch of a long capillaries which is combined between themselves due to action of the surface tension forces so it is easy to estimate a limit value of the capillary diameters $d \leq 3 \cdot 10^{-2}$ cm, that corresponds to (7).

## **5. Conclusion.**

It is necessary to pay attention a physical sensation of presence the SSO into the sea water is such as if you will immerse your hand into the water, because here presences only the STF. What is happened with the SSO after that, as a storm is ended? A part of its immerses on some depth and it will be not prove themselves nothing before the next storm, which will be able to extract its on the OS. Another part of its and in first of all the HFC will insignificantly to tower above the OS. Rising above the OS its parts dries up (due to action of the Sun) and loses of a connecting their fluid films and disintegrates on composing fragments, which, in one's turn, will to float or immerse depending on its individual buoyancy. Such a floating fragments of the discussing OSS have tendency to forming the OSS, which corresponds of the OS rough level at the present moment of time and in a given region. Evidently, a time of formation more or less a stable and homogeneous net of such structure should be individual and finite quantity for each of the OS rough level. Therefore it is clear, why tracks of a ship routes (by means of observation with cosmos orbit) at a calm it is nice views up to half thousand kilometers, that is during of 12 hours, because a reflective ability of the OS depends on a presence, form and typical scales of the OSS. Indeed, a screw propeller of ships destroys and grinds of the OSS disturbing the homogeneity of formed the OSS, and time of its reconstruction demands of a wide time interval and identity of history of its establishment. That is, *individuality of the OSS in the given region determines by the all prehistory of its formation.*

## Acknowledgements.

The author is deeply grateful to A.B. Kukushkin for a decade long collaboration. Special thanks to V.I. Kogan for invariable support and interest to a research of skeletal structures.